\documentclass{emulateapj}
\pdfoutput=1
\usepackage{graphics}

\newcommand{\beq}	{\begin{equation}}
\newcommand{\eeq}	{\end{equation}}
\newcommand{\beqa}	{\begin{eqnarray}}
\newcommand{\eeqa}	{\end{eqnarray}}
\newcommand{\calm}	{{\cal M}}

\newcommand{\vecB}	{{\bf B}}

\newcommand{\vecnabla}	{{\bf\nabla}}

\newcommand{\cross}	{{\bf \times}}
\newcommand{\gad}       {{\gamma_{\rm AD}}}

\newcommand{\alfven}    {{Alfv$\acute{\rm e}$n }}

\newcommand{\alfvenic}  {{Alfv$\acute{\rm e}$nic }}

\newcommand{\chio}	{\chi_{i0}}

\newcommand{\mai}	{{\calm_{{\rm A}i}}}

\newcommand{\rad}	{R_{\rm AD}}

\newcommand{\avg}[1]    {{\langle #1 \rangle}} 

\newcommand\brms        {B_{\rm rms}}

\newcommand{\e}	        {$^{-1}$}

\newcommand{\eee}	{$^{-3}$}

\newcommand{\ma}	{{\calm_{\rm A}}}

\newcommand{\radl}	{R_{\rm AD}(\ell)}

\newcommand{\acf}	{a_{\rm CF}}

\newcommand{\radlo}     {R_{\rm AD}(\ell_0)}

\shorttitle{MHD Turbulence Simulations with Ambipolar Diffusion}
\shortauthors{Li, McKee, \& Klein}
\begin{document}
\title{Sub-\alfvenic Non-Ideal MHD Turbulence Simulations with Ambipolar Diffusion: III. 
Implications for Observations and Turbulent Enhancement}
\author{Pak Shing Li}
\affil{Astronomy Department, University of California,
    Berkeley, CA 94720}
\email{psli@astron.berkeley.edu}
\author{Christopher F. McKee}
\affil{Physics Department and Astronomy Department, University of California,
    Berkeley, CA 94720}
\email{cmckee@astro.berkeley.edu}
\and
\author{Richard I. Klein}
\affil{Astronomy Department, University of California,
    Berkeley, CA 94720; and Lawrence Livermore National Laboratory,\\
    P.O.Box 808, L-23, Livermore, CA 94550}
\email{klein@astron.berkeley.edu}

\begin{abstract}
Ambipolar diffusion (AD) is believed to be a crucial process for redistributing magnetic flux in 
the dense molecular gas that occurs 
in regions of star formation. We carry out numerical simulations of this process in regions of low ionization using the heavy ion approximation. The simulations are
for regions of strong field (plasma $\beta=0.1$) and mildly supersonic turbulence
($\calm=3$, corresponding to an \alfven\ mach number of 0.67). The velocity power spectrum of the neutral gas changes from an Iroshnikov-Kraichnan spectrum in the case of ideal MHD to
a Burgers spectrum in the case of a shock-dominated hydrodynamic system.  
The magnetic power spectrum shows a similar behavior.
We use a 1D radiative transfer code to post-process our simulation results; the
simulated emission from the CS $J=2-1$ and H$^{13}$CO$^+\ J=1-0$ lines shows that the effects of AD are observable in principle.
Linewidths of ions are observed to be less than those of neutrals, and we
confirm previous suggestions that this is due to AD. We show that AD is unlikely to affect
the Chandrasekhar-Fermi method for inferring field strengths unless the AD is stronger
than generally observed. Finally, we present the first
fully 3D study of the enhancement of
AD by turbulence, 
finding that AD is accelerated by factor 2-4.5 for non self-gravitating systems
with the level of turbulence we consider.

\end{abstract}

\section{Introduction}

The magnetic flux problem is one of the classic problems in star formation: How is it
that stars form with magnetic fluxes many orders of magnitude less than the gas out of
which they originated? Ambipolar diffusion (AD), the process by which the neutral component
of a magnetized plasma can move relative to the ion component, was identified as a crucial
mechanism for resolving this problem more than half a century ago \citep{mes56}.
Subsequent study of the quasi-static AD-driven collapse of an isolated cloud core \citep[e.g.][]{spi68,mou76,mou77,mou79,nak78,shu83,fie92,fie93} showed that AD could indeed reduce the flux by a large factor.  However, both observations \citep[e.g.][]{zuce74,zucp74} and theory \citep[e.g.][]{aro75,mac04,bal07,mck07} have revealed that supersonic turbulence in molecular clouds plays a central role on the whole star formation process. Since molecular clouds are  magnetized \citep[e.g.][]{cru99,hei05,tro08}, we are confronted with the problem of understanding AD in a turbulent medium.

While important insights have been obtained through analytic studies of AD in a turbulent medium in two dimensions \citep{zwe02,fat02,kim02,hei04}, the only feasible approach for addressing this problem in three dimensions is numerical simulation.  Unfortunately, even numerical simulation faces difficulties in the study of turbulent, magnetized  molecular clouds with AD: If the neutrals and ions are treated as a single fluid, then AD becomes a nonlinear diffusion problem with a time step that scales as the square of the grid spacing, which becomes increasingly challenging at high resolution. On the other hand, if the neutrals and ions are treated as separate fluids, then the time step scales inversely as the fastest signal velocity.  Observations \citep[e.g.][]{cas98,ber99} show that the ionization fraction in star-forming regions of molecular clouds is typically $\la 10^{-7}$.  For such low ionization fractions, the ion \alfven speed is at least three orders of magnitude larger than the neutral \alfven velocity, 
which leads to a time step that is smaller than in the hydrodynamic case by the same factor.  A two-fluid simulation of AD in a star-forming region would require a time step $\ga 10^3$ times smaller than for an ideal MHD simulation; on a $512^3$ grid, a simulation lasting several dynamical times would require $\sim 10^7$ CPU hours, even on the most advanced parallel computers using the most efficient massively parallelized MHD codes.

\citet{li06} developed the Heavy-Ion Approximation for MHD with AD and tested it with the ZEUS-MPAD code (this approximation was used independently by \citet{ois06}).  The Heavy-Ion Approximation reduces the ion \alfven\ velocity by increasing the mass of the ions, while at the same time reducing the coupling to the neutrals so as to maintain the same rate of momentum transfer between the two species. This approximation is valid only when the ion inertia is negligible. 
As pointed out by \citet{til08}, it is therefore impossible to represent high-frequency MHD waves that propagate in the ion fluid with this approximation. 
However, this is a positive feature of our method, not a flaw: it enables
simulations of very weakly ionized plasmas, such as those
in molecular clouds, to proceed {\it without}
having to follow waves with phase velocities that can reach thousands of km s\e\ in a fluid with velocities of only km s\e.  Contrary to the statement by \citet{li10} that ``ion-neutral decoupling does not appear in the two-fluid treatment with heavy ion approximation," our series of papers on turbulent AD clearly demonstrates that ion-neutral decoupling clearly appears in two-fluid treatment when the Heavy Ion Approximation is used correctly.

Through a series of $256^3$ simulations, \citet{li08} (hereafter Paper I) showed that this approximation can successfully accelerate sub-\alfvenic turbulence simulations with AD by a factor of 100. They studied the statistics of non-ideal MHD turbulence with AD using two fluids, neutrals and ions.  The purpose of Paper I was twofold: (1) to study the convergence behavior of the Heavy Ion Approximation in order to determine the optimal increase in the ion mass; and (2) to investigate the effect of AD on the statistical properties of a supersonically turbulent system.  The results were presented as a function of the AD Reynolds number, which is the ratio of the characteristic AD timescale to the flow time, or, equivalently, the ratio of the size of the system to the characteristic AD lengthscale \citep{zwe97},
\beq
\radlo\equiv\frac{4\pi\gad\bar\rho_i\bar\rho_n \ell_0 v}{\brms^2},
\label{eq:radell}
\eeq
where $\gad$ is the ion-neutral coupling parameter, $\bar\rho_{i,\, n}$ is the mean density of the ions and neutrals, respectively, $\ell_0$ is the size of the system (for the simulation, it is the size of the turbulent box), $v$ is the rms velocity, and $\brms$ is the rms magnetic field. 
The importance of AD increases as $\rad$ decreases. In Paper I we found that the power spectra of the velocity and the magnetic field change from those for ideal MHD to those for hydrodynamics as $\radlo$ goes from very large values to values less than unity. 
Most of the simulations in Paper I used a unigrid $256^3$ in size, with only one $512^3$ model to show that the results were converged.

Since stars form from dense clumps, the properties of clumps in a turbulent medium are important.  To study dense clumps that arise in supersonic turbulence in the presence of AD, we carried out four more $512^3$ models with different values of $\radlo$ in \citet{mck10} (hereafter Paper II).  The AD models presented in Paper II span four orders of magnitude in $\radlo$. It should be noted that neither Paper I nor Paper II included gravity.
In paper II, we used the CLUMPFIND algorithm \citep{wil94} to determine the clumps in our simulations.  We defined "clumps" as connected regions with a density larger than the mean density of the turbulent box (see paper II on details of how we used CLUMPFIND to extract clumps from the simulations).
We found that AD affected the clump mass function and mass-to-flux ratio, even in the absence of gravity.  We introduced the categorization of AD into five regimes by comparing the gas flow time, $t_f=\ell_0/v$, with the ion-neutral and neutral-ion collision times, $t_{\rm in}$ and $t_{\rm ni}$, respectively.  The typical value of $\radlo$ for 15 cloud clumps with measured magnetic fields \citep{cru99} was found to be $\sim 20$, in excellent agreement with the theoretically expected value for self-gravitating clumps. Paper II also presented a detailed discussion of scaling relations in simulations with AD and showed that simulations that include self-gravity cannot vary the self-gravity and AD independently unless the ionization is treated as a free parameter.  The results of Papers I and II  show that AD can have important effects on the clumps even before the core collapse phase that is the focus of most quasi-static AD-driven star formation models.

In work that is complementary to that in the present series of papers, \citet{kud08,kud11} have simulated MHD turbulence, including AD, in a thin, self-gravitating slab of gas.  The turbulence was initialized in the plane of the slab and allowed to decay. The focus of their work was on the time required for bound cores to form in gas in which the magnetic field initially dominated gravity. By contrast, our work considers the effects of AD on MHD turbulence in the absence of gravity. The turbulence is driven continually in all three dimensions so as to maintain a constant Mach number.

In this paper, we present our results on the observational implications of AD in turbulent molecular clouds.  In \S2, we summarize our models and the assumptions that we have adopted.  In \S3, we use our higher resolution models to revisit the power spectra of velocity and magnetic field found in Paper I. 
In \S4, we study the linewidth-size relation of the two fluids using different statistical methods and compare with other work using ideal MHD simulations.  \S5 discusses the ratio of the line widths of the ions and neutrals, which was suggested as an indicator of AD by \cite{hou00a,hou00b}. One of the challenges facing observers studying molecular clouds is determining the magnetic field strength inside molecular clouds. A commonly used method for doing this is the Chandrasekhar-Fermi method (CF method) \citet{cha53}, and we discuss the effects of AD on this method in \S6.
Turbulent enhancement of diffusion processes is commonly observed in experiments, observations, and numerical simulations, and in \S7 we discuss the turbulent enhancement of AD in molecular clouds.
We summarize our findings in \S8.

\section{Models and Assumptions}

The results presented and discussed in this paper are from the same models presented in Paper II.  There are a total five AD models with $\rad$ incremented by a factor of 10 from 0.12 to 1200 using a $512^3$ grid.  Simulations of an ideal MHD model, corresponding to $\rad = \infty$, and a pure hydro model, corresponding to $\rad = 0$ are also included for comparison (see Table 1).  All models are driven by a fixed pattern turbulence field, generated by the \citet{mac99} recipe, to maintain a constant Mach number $\calm = 3$.  The turbulence is driven at the largest scale between wavenumbers $1 \leq k\leq 2$.  (The dimensionless wavenumber $k$ is related
to the physical wave number by $k=k_{\rm phys}\ell_0/(2\pi))$.  
The inertial range of the turbulence extends from $k\sim 3$ to $k\sim 20$, where the upper limit is set by numerical diffusion.
Boundaries are periodic and the turbulent region is initially threaded by a uniform magnetic field with a plasma $\beta = 0.1$.  The AD models are based on the assumption of ion conservation. 
Discussion of this assumption, including comparison with that of ionization equilibrium, can be found in the Appendix of Paper I and in Paper II.  
For our AD models, there is no clear difference between using ion conservation and using ionization equilibrium. 
Gravity is not included in any of the simulations.

Molecular gas is observed to be approximately isothermal, and we assume that here. In fact, temperature fluctuations are expected due to the intermittency of turbulence; in supersonic turbulence, gas is heated in shocks, for example. Since the cooling lines in molecular gas are often optically thick \citep{gol78}, accurate calculation of the temperature is a complex problem in radiative transfer that is beyond the scope of this paper. We note that \citet{pan09} have shown that
cosmic ray heating dominates in most of the volume of a supersonically turbulent medium such as we consider, and the isothermal assumption holds there. AD heating dominates in a small fraction of the gas, however, and that will be the subject of a future paper.

In Paper II, we gave a detailed discussion on how to convert the results of simulations to physical units for easy comparison to observations, and we do not repeat that here.  As a reminder, for the AD model with $\rad = 12$, the size of the simulation region is 0.41 pc, the total mass is $92 M_\sun$, the turbulent flow time across the region is 0.71 Myr, and the initial mean magnetic field is $90 \,\mu G$.

\section{High Resolution Simulation Power Spectra and the Change of Cloud Morphology}
\label{sec:power}

\begin{figure}
\epsscale{.80}
\plotone{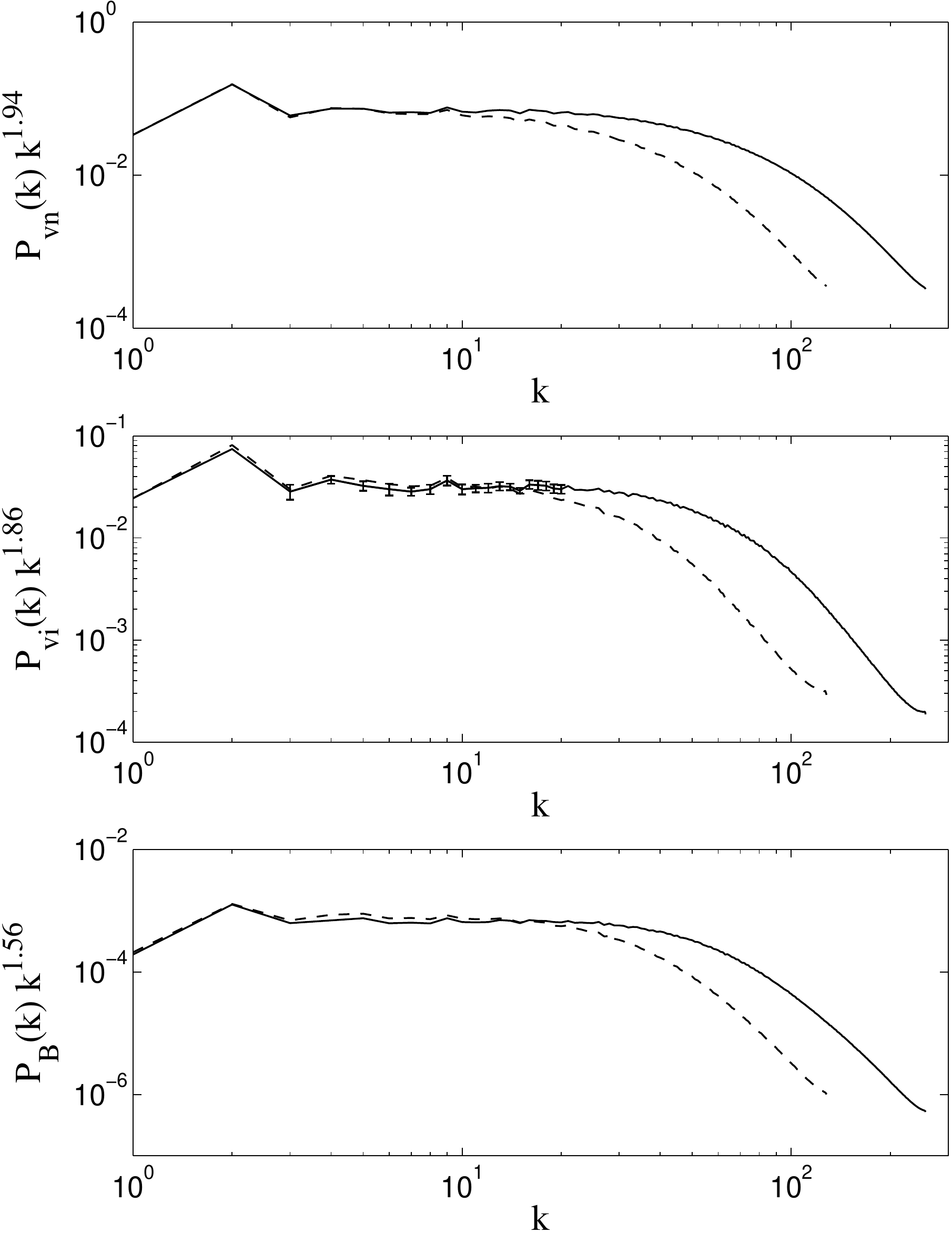}
\caption{Compensated velocity and magnetic field power spectra of $256^3$ (dashed line, model m3c2 in Paper I) and $512^3$ models with $\radlo = 1.2$ (solid line, model m3c2r0 in this paper).  See Table 3 in Paper I and Table 2 in this paper for the compensated power indexes.  The inertial range in $512^3$ resolution extends up to $\sim k=20$.
The error bars of the inertial range are plotted in the middle panel to show the typical sizes of the error bars in the data fitting.
\label{fig1}}
\end{figure}

In Paper I, we discussed the turbulent power spectral indexes as functions of $\rad$ based on $256^3$ simulations, although we presented one $512^3$ model as part of a convergence study.  We considered a range of $\radlo$ from 0.12 to 1200.  With the doubled resolution of the simulations in this paper, the inertial range is well developed up to at least $k$ = 20 (Figure \ref{fig1}).   The resulting spectral indexes for the five models using $\chi^2$ fitting from $k$ = 3 to 20 are given in Table 2.
All the fitting results and the plots shown in this paper are time-averaged results from 14 data dumps over two crossing times unless specified otherwise.
The spectral index, $n_x$, for a variable $x$ is defined by 
\beq
P(k)dk\propto k^{-n_x}dk, 
\eeq
where $P(k)dk$ is the power in the wavelength interval $dk$.
The uncertainties in the power-law fits to the inertial range slopes are improved.  Comparison of the results listed in Table 3 of Paper I with the new, higher-resolution models shows that most of the results from the $256^3$ models are within the uncertainties.  This confirms the conclusions in Paper I on how the power-law indexes of velocity and magnetic field change from ideal MHD at very large $\radlo$ to pure hydrodynamics at small $\radlo$, where ambipolar diffusion dominates.  We plot the indexes of the ion and neutral velocity power spectra and of the magnetic field power spectra in Figure \ref{fig2} for better visualization of the transition of the turbulent systems.  At large $\rad$, both the ion and neutral velocity spectra are close to the Iroshnikov-Kraichnan \citep{iro63,kra65} spectrum, which has $n_v=3/2$.  As $\rad$ decreases, the neutral velocity spectrum evolves towards the shock-dominated Burgers spectrum \citep{bur74}, which has $n_v=2$ and is expected for purely hydrodynamic supersonic turbulence.  Note that for supersonic, super-\alfvenic turbulence, the velocity spectral index is also similar to a Burgers spectrum, with $n_{vn} = 1.9 \sim 2.0$ \citep{pad07}.  The spectral index for the ion velocity also becomes steeper but does not appear to be a Burgers spectrum yet, with $n_{vi} \sim 1.85$ over the range $\radlo \sim 0.1-10$.  Apparently, the infrequent collisions with neutrals at such small values of $\rad$ are not able to turn the ion velocity spectrum into a Burgers spectrum.
The transition between ideal MHD and the AD-dominated, quasi-hydrodynamic behavior
of the turbulence occurs over the range $\radlo \sim 1-100$, i.e., at relatively large AD Reynolds numbers. In the recent study by \citet{loo08} on the propagation of nonlinear MHD waves in two dimensions, the effects of ambipolar diffusion also set in at significantly larger scales than one would naively expect at around $\radl \sim 1$.

In Paper II, we categorized the effect of AD into five regimes:

\begin{itemize}
\item[I.] Ideal MHD ($\rad\rightarrow\infty$): The ions and neutrals are
perfectly coupled.

\item[II.] Standard AD ($t_f>t_{ni}\gg t_{in}$, corresponding to $\rad>\ma^2$): 
The neutrals and ions are coupled together over a flow time.

\item[III.] Strong AD ($t_{ni}>t_f>t_{in}$, corresponding to $\ma^2>\rad>\mai^2$):
The neutrals are no longer coupled to the ions in a flow time, but the ions
remain coupled to the neutrals.

\item[IV.] Weakly coupled ($t_{in}>t_f$, corresponding to $\mai^2>\rad$): 
The ions and neutrals are only weakly coupled and act almost independently.

\item[V.] Hydrodynamics ($\rad\rightarrow 0$): The neutrals are not affected by
the trace ions and act purely hydrodynamically.

\end{itemize}
\noindent
Here $t_f$ is the flow time, or the dynamical time scale of the turbulence, $t_{ni}$ is the neutral-ion collision time scale, and $t_{in}$ is the ion-neutral collision time scale.  From Table 2 and Figure \ref{fig2}, the transition between ideal MHD and the AD-dominated, quasi-hydrodynamic behavior of the turbulence occurs in the Standard AD regime.  Near the beginning of the Strong AD regime (i.e., for $\rad$ somewhat less than $\ma^2$), the neutral power spectrum is very much the same as a Burgers spectrum.   Fifteen molecular clumps with measured magnetic fields \citep{cru99} have $\rad$ from 3 to $\sim 70$, with a logarithmic mean of the AD Reynolds number $\sim 20$ (see Paper II).  All these clumps are inside the Standard AD regime, and measurements of their velocity power spectra should show a transition from a spectrum close to the Iroshnikov-Kraichnan spectrum to one close to a Burgers spectrum as $\radlo$ goes from the highest observed value to the lowest.  The power index of the magnetic field goes through a similar steepening as AD becomes stronger in regime II, mimicking closely the transition in the velocity power indexes.

Previous studies of the density power spectrum for supersonic hydrodynamic and MHD turbulence \citep[e.g.][]{kim05,kow07} have found that the density spectral index, $n_\rho$, is dependent on the sonic Mach number but (for MHD turbulence) not very sensitive to the \alfven Mach number.  Using $512^3$ supersonic hydrodynamic simulations, \citet{kim05} reported that $n_\rho$ varies from 0.52 to 1.73 as the Mach number varies from 12 to 1.2.  In the limit of high Mach number, shocks compress the gas into infinitely thin sheets, leading to $P_{\rho} (k) \propto k^0$; pressure forces become important as the Mach number decreases, leading to $n_\rho\sim 2$ at $\calm\sim 1$ \citep{sai96}.  From $256^3$ MHD supersonic turbulence simulations, \citet{kow07} found that $n_\rho$ is shallower than in hydrodynamic supersonic turbulence: For $\ma \sim 0.7$, they found $n_\rho = 0.5\pm0.1$ at a sonic Mach number $\calm = 7.0\pm0.3$ and $n_\rho=1.3\pm0.2$ at $\calm = 2.2\pm0.03$.

The density spectral indexes for all our models are tabulated in Table 2.  The \alfven Mach number of all the turbulence models reported in this paper is $\ma=0.67$ and the sonic Mach number is $\calm = 3$.  First we check to see if our models at the extremes of ideal MHD and pure hydrodynamic turbulence are consistent with other simulation results.  From the results of \citet{kow07} and \citet{kim05}, we expect $n_\rho \sim 1.17$ and 1.2, respectively, for ideal MHD (model m3i) using a simple linear interpolation; consistent with this, we find $n_\rho = 1.08\pm0.07$.  Our AD model m3c2r-1, with $\radlo = 0.12$, should be very close to the hydrodynamic limit; we find $n_\rho \sim 1.23\pm0.05$, consistent with the hydrodynamic results of \citet{kim05}.  The fact that $n_\rho$ and $n_{\rho,n}$ for models m3i and m3c2r-1 are numerically about the same is a coincidence for our particular choice of sonic and \alfven Mach numbers.  However, the amplitudes of the density power spectra for the m3i and m3c2r-1 models are not the same because AD allows larger density contrasts (larger dispersion in the density PDF, see Paper I) to occur in the neutral gas for model m3c2r-1 ($\radlo=0.12$) than in the gas for model m3i (ideal MHD).  This is most likely because compression in the ideal MHD case is primarily along the field lines, whereas for low $\radlo$ the compression can occur in all three dimensions. 

From Table 2, we see that AD has an interesting effect on the density power spectrum as $\radlo$ varies, even though the values of $n_\rho$ for the ideal MHD and pure hydro regimes are roughly the same for our choice of turbulence initial conditions.
The index $n_\rho$ steepens from 1.08 to 1.77 as $\radlo$ drops from $\infty$ to 12 and then changes back to 1.23 at $\radlo = 0.12$.  
We attribute the steepening between ideal MHD and $\radlo=12$ to the fact that
at high values of $\radlo$, AD is most effective on small scales, and this leads to
damping of the density fluctuations on such scales. As $\radlo$ decreases further, AD becomes effective on all scales and the results approach the hydrodynamic limit. 
These results are shown graphically in
Figure \ref{fig3}, where we plot the density power spectrum of model m3i and the neutral density power spectra of models m3c2r2, m3c2r1, and m3c2r-1. The neutral density power spectrum of model m3c2r3 ($\radlo=1200$) is similar to the density spectrum of m3i, and that of model m3c2r0 ($\radlo=1.2$) is similar to that of m3c2r-1, so they are not plotted.  The range $\radlo = \infty$ to 12 includes the transition from AD regime I to AD regime II.  

AD also has a significant effect on the morphology of the gas.
In Figure \ref{fig4}, we plot density slices from models m3c2r3 ($\radlo=1200$), m3c2r1 ($\radlo=12$), and m3c2r-1 ($\radlo=0.12$) in the $x-z$ plane at the same value of $y$; the mean field is in the $z$-direction.
The density slices are at taken at the end of the simulations and 
displayed on a linear scale in order to better show the changes in the spatial distribution of the high density gas.  Figure \ref{fig4}a shows clear filamentary structures aligned roughly along the mean $B$-field direction in model m3c2r3.  The curved filamentary structures trace the perturbed $B$-field lines.  In this model, gas can freely move along the $B$-field but cannot easily cross it.  When AD becomes important, the morphology changes because of diffusion across the $B$-field (see Figure \ref{fig4}b, $\radlo=12$).  The density spectral index steepens as discussed above and the maximum density also increases.  In going from $\radlo = 12$ to 0.12, the system moves from AD regime II into regime III, and the morphology changes again (see Figure \ref{fig4}c).  Fragmentation becomes important because the magnetic field can no longer suppress motions normal to the mean field direction, and as a result, high-density structures form at much smaller scales and the density power spectrum flattens.  The increase in fragmentation from $\radlo = 12$ to 0.12 is also reflected in the clump mass function, size, and geometry, as discussed in Paper II.  The changes in the density power spectrum as AD becomes more important are thus mirrored in the morphology of the gas.

Figure \ref{fig5} shows what observers would see if they observed the gas in our simulations with the Atacama Large Millimeter Array (ALMA).  We used the CASA simdata2 software to create the simulated observations.  CASA (Common Astronomy Software Application) is a suite of software tools for calibration and analysis of radio astronomical data \citep{jae08}.  Simdata2 turns a model of the sky into the visibilities that would be measured with ALMA and then produces a synthetic image from the model visibilities.  The panels on the right column are simulated images 
from models m3c2r3, m3c2r1, and m3c2r-1 assuming a 2-hour observation of CS $J=2-1$ emission at a frequency of 98 GHz.  The panels on the left are simulated images of the corresponding ion component from the simulation,
H$^{13}$CO$^+\ J=1-0$ emission at frequency of 86.6 GHz, also for 2-hour observation.  The viewing direction is normal to the mean field direction, which is along the vertical axis.  We use the 1D radiative transfer code "SimLine" \citep{oss01} to compute the line emission from a region of a molecular cloud with size 0.41 pc and density $n$(H) = $10^4$ cm$^{-3}$.  Since $\rad$ is proportional to the ionization fraction, the the two limiting cases m3c2r3 and m3c2r-1, with $\rad = 1200$ and 0.12 respectively, are models with the highest and lowest ionization fractions.  The abundances of some molecules, such as CS and HCO$^+$, are found to be insensitive to high ionization environments, both from observations and modeling, even over a two-order-of-magnitude change in the ionization rate \citep{far94,lep96,mei11}.  Modeling of the abundances of CS and HCO$^+$ shows that there could even be an anti-correlation between abundance and ionization rate at high column density.  For simplicity, we fix the CS abundance to be $10^{-9}$ for all models \citep[e.g.][]{fur11}.  
We assume the mean abundance HCO$^+$ relative to H$_2$ is $2 \times 10^{-9}$ for the standard ionization model, m3c2r1 \citep[e.g.][]{pad04,fur11}; this model has $\rad = 12$, close to the median value in Crutcher's sample of molecular cloud clumps \citep[][and see paper II]{cru99}. 
For the low-ionization model, mc3cr-1, we assume that the mean abundance of HCO$^+$ tracks the total ionization, which is 100 times less.  For the high-ionization model, m3c2r3, we adopt a mean abundance for HCO$^+$ of $5 \times 10^{-9}$.  We assume that the abundance of H$^{13}$CO$^+$) is 2\% of that for HCO$^+$ \citep[e.g.][]{pur06,fur11}.  
Since the ion density varies at different location in the turbulent box, the local abundance of H$^{13}$CO$^+$ relative to H$_2$ will depend on the local ion density from our simulations.  The relevant molecular data are taken from the JPL catalogue \citep{pic98} and the CDMS database \citep{mul01}.   We compute the line emission intensities along the line of sight as functions of frequency from these models using SimLine.  Both lines are mostly optically thin and have maximum optical depths of order unity.  The results are used as input for CASA simdata2 to generate the images that would be observed with ALMA.  The size of the turbulent box is assumed to be $1\arcmin$ in the maps.  When AD is negligible, as in the case of model m3c2r3, the simulated images of the ion and neutral gases are basically identical, as expected.  When AD becomes important, the simulated images show that the ion and neutral emission differ from each other.  The simulated images in Figure \ref{fig5} demonstrate that decoupling of neutrals from ions is observable in principle at values of $\rad$ typical of clumps with measurable magnetic fields.  Observation of a difference between ion and neutral images is not in itself sufficient to conclude that AD is operating, however, since the difference could be due to chemical effects as well.; if the images are the same, however, AD is unlikely to be important.

\section{Linewidth-Size Relation}
\label{sec:lws}
Larson (1981) was the first to point out the correlation of size and velocity dispersion in molecular clouds, which he attributed to turbulence:
\beq
\sigma_v \propto R^q.
\label{eq:lwseqn}
\eeq
He found $q = 0.38$, but subsequent observations using various techniques \citep[e.g.,][]{sol87,mie94,laz04,oss06} have found $q \simeq 0.5$.  
Since
\beq
\sigma_v(\ell)^2 = \int_{k=\ell_0/\ell}^{\infty}{P_v(k) dk},
\label{eq:intpower}
\eeq
the spectral index of the velocity power spectrum, $n_v$, is related to the linewidth-size parameter $q$ by 
\beq
n_v=2q+1.
\label{eq:q}
\eeq
\citet{pas88} pointed out that with $q = 0.5$, the linewidth-size relation corresponds to the Burgers spectrum, $P_v(k) = k^{-2}$.  

Note that the linewidth-size index $q$ in equation (\ref{eq:lwseqn}) is not identical to the one inferred by observers from a Principal Components Analysis (PCA: \citealp{hey97,bru02}), who write the linewidth-size relation as 
\beq
\sigma_{v,\,\rm obs} \propto R^\alpha.
\eeq
As discussed by \citet{bru02}, the information retrieved from a principal component analysis of the observed spectra involves a line-of-sight projection of the emissivity (modulated by the opacity for optically thick lines), which depends on the velocity, density, temperature, and abundance, all of which vary along the line of sight.  (By contrast, the data analyzed by \citet{lar81} and \citet{sol87}, for example, focused primarily on entire clouds, where the sizes are determined by a transition from molecular to atomic C.) \citet{bru02} use numerical models to find a conversion between the observed linewidth-size index $\alpha$ and the velocity power spectral index $q$.
For example, for a log-normal density distribution, they find that an observed value of $\alpha\simeq 0.6$ corresponds to $q\simeq 0.5$.

Here we compare the linewidth-size relation determined using the power spectrum in Fourier space with that in physical space.  We shall not discuss
the alternative method of determining the velocity power spectral index from
spectral maps of emission lines  \citep{laz00}, which has been verified by high resolution simulations \citep{pad06}.

In the box-decomposition method
\citep[e.g.][]{lem09},
we divide the turbulent box into successively smaller boxes of equal size, $\ell = \ell_0/2^m$, until the boxes are of size $2 \times 2 \times 2$ grid cells.  We use volume weighting to compute the velocity dispersions of the ions and neutrals in the local frame of each box so that we can compute the velocity power spectral index directly.  \citet{lem09} also used the volume-weighted velocity dispersion in their box decomposition analysis,
but \cite{fal10} used the density-weighted velocity, $\rho v$, in their analysis.

In Figure \ref{fig6}, we plot the ion and neutral velocity dispersions of models m3c2r-1 ($\radlo=0.12$), m3c2r1 ($\radlo=12$), and m3i (ideal MHD) as functions of length scale based on the box decomposition method.
Note that in model m3c2r-1, which has strong AD, the ion velocity dispersion is smaller than the neutral velocity dispersion at all length scales.  For model m3c2r1, the ion velocity dispersion is just slightly smaller than the neutral velocity dispersion (see Section \ref{sec:inlr} for further discussion of the ion-neutral line ratios).  The ion and neutral velocity dispersions for model m3c2r3 are nearly the same as those for the ideal MHD model, m3i.

A well defined power-law region can be identified in the linewidth-size relation plotted in Figure \ref{fig6},.  This power-law region almost exactly overlaps the inertial range in the velocity power spectrum, $k \simeq 3 - 20$.  The time-averaged power indexes in the inertial range based on the box decomposition method, $q_{\rm bd}$, are listed in Table 3.  These indexes are derived from a $\chi^2$ fit of the three data points corresponding to $\ell/\ell_0=1/4,\,1/8,$ and 1/16.  This linewidth-size parameter in physical space, $q_{\rm bd}$, can be compared to the one implied by the velocity power spectrum in Fourier space, 
$q_{\rm ps} = (n_v-1)/2$.  The values of $q_{\rm ps}$ for the ions and neutrals, $q_{\rm i,ips}$ and $q_{\rm n,ips}$, are also listed in Table 3.  We can immediately see that there is a large discrepancy between $q_{\rm ps}$ and $q_{\rm bd}$.  The reason is simple: $q_{\rm  ps}$ is based on only the inertial range, whereas each box that goes into the determination of $q_{\rm bd}$ includes all scales smaller than the box size; as shown in Figure \ref{fig6}, dissipation reduces the velocity dispersion on scales below the inertial range.  Therefore, to obtain a value of $q$ from the power spectrum that is comparable to that from the box decomposition method, we need to do an integration from the actual velocity power spectrum, as in equation (\ref{eq:intpower}); we term the resulting power-law index $q_{\rm dps}$ since it includes the dissipative range.  
Note that the fit for $q_{\rm dps}$ is based on using the same three values of $\ell/\ell_0$ in equation (3) as were used in determining $q_{\rm bd}$.  Because of the much lower power at large $k$ due to numerical dissipation, $q_{\rm dps}$ is steeper than $q_{\rm ps}$, which is estimated from just the inertial range.  The results for both ions and neutrals are listed in Table 3, which shows that $q_{\rm dps}$ matches $q_{\rm bd}$ much better than $q_{\rm ps}$ does.  From this exercise, we can see that it is inappropriate to compare the value of $q$ determined using box decomposition with observation because the uncontrollable numerical dissipation region at large $k$ (which some codes confound with a large bottleneck effect) significantly affects the linewidth-size index.

\section{Ion-Neutral Linewidth Ratio}
\label{sec:inlr}
Observations of lines of molecular ions (e.g. HCO$^+$, HCS$^+$) and of neutral molecules (e.g. HCN, H$^{13}$CN) show that in many cases the neutral linewidths exceed the ion linewidths \citep {hou00a,hou00b,hou02}.  This is a striking result, since HCO$^+$ has a lower critical density than HCN and therefore should be more spatially extended and correspondingly have a larger linewidth \citep{hou00a}.  These authors suggested that AD is responsible for this effect and presented a microphysical model to account for it. They began by assuming that they were in the frame in which the electric field vanishes.  For molecular clouds, in which the ion gyrofrequency is large compared to the collision frequency, this corresponds to the ion frame. They showed that if the neutrals drift relative to the ions, the ions will acquire motions perpendicular to the field that correspond to the temperature associated with the neutral drift velocity, $m_i\sigma_i^2\simeq m_n v_{\rm drift}^2$. In fact, this ion heating is well known in the theory of C-shocks \citep{dra80}. \citet{hou00a} then concluded that because the velocity dispersion of the hot ions is less than the drift velocity of the neutrals (since the ion mass is much larger than the mean neutral mass), the ion linewidths should be narrower than those of the neutral molecules. This argument completely neglects the turbulent motions of the ions (which are associated with hydromagnetic waves in the directions normal to the magnetic field), however, and so is invalid. Subsequently, \citet{lih08} 
suggested a different microphysical model for the effect of AD on the linewidths: they pointed out that the ambipolar drift velocities needed to account for the observations would occur only on small scales, such that $\rad\la 1$.  They assumed that the turbulent velocities of the ions would damp out on scales smaller than this, whereas those of the neutrals would persist down to the neutral viscous scale; as a result, the net velocity dispersion of the ions would be less than that of the neutrals.  
\citet{li10} showed that the ion-neutral linewidth ratio was smaller in regions of strong magnetic field, which is consistent with AD since $\radlo\propto B^{-2}$.  \citet{hez10} assumed that the damping
of the ion turbulence would occur at a well-defined length scale corresponding to 
$\rad=1$ and inferred the value of that length scale in DR21(OH).
On the theoretical side, \cite{fal10}  presented arguments supporting the conjecture that AD is responsible for the small observed values of the
ion-neutral linewidth ratio, 
although their simulations, being ideal MHD, could not directly
test the conjecture. In simulations that are closest to the ones presented here,
\citet{til10} carried out simulations with a two fluid code and showed that the ion-neutral
linewidth ratio systematically decreased as $\radlo$ decreased, over a similar range as
we consider. They did not use the Heavy-Ion Approximation, and as a result their
resolution was limited to $192^3$.

Our results are fully consistent with the conjecture by \citet{hou00a} that AD is responsible for the observation that ion linewidths are narrower than neutral ones.
However, our results do not show the abrupt decline in the ion velocity power spectrum that they postulated, so they do not support the more detailed model put forth by \citet{lih08}.  It must be borne in mind that since our simulations are isothermal, we do not include the drift heating of the ions, and as a result our ion-neutral linewidth ratios are lower limits.

In Figure \ref{fig6}, we have already seen that the mean velocity dispersion of the ion component, $\sigma_{v,i}$, is $always$ less than or equal to the mean velocity dispersion of the neutral component, $\sigma_{v,n}$, at all length scales.  The stronger the AD effect (smaller $\rad$), the smaller $\sigma_{v,i}$ is compared to $\sigma_{v,n}$.  We can plot the ion and neutral velocity dispersions at different length scales using box decomposition in the same way as in Figure 3 of \citet{hou02} for comparison.  In Figure \ref{fig7}a, we plot the volume-weighted mean $\sigma_{v,i}$ versus the volume-weighted mean $\sigma_{v,n}$ computed using box decomposition for models m3c2r3, m3c2r1, and m3c2r-1, corresponding to $\radlo=1200,\;12,$ and 0.12, respectively.  The error bars show the errors of the time-averaged mean. The velocity dispersions from the different length scales are plotted: the highest velocity dispersions are from the largest boxes ($\ell_0/4$), and the lowest are from the smallest ($\ell_0/128$).  The highest three points are inside the inertial range.  We can see that the data points from length scales smaller than the inertial range also follow the same trend as those inside the inertial 
range. This result is consistent with the observed linewidth ratios reported in \cite{li10}.  The solid line corresponds to $\sigma_{v,i}/\sigma_{v,n} = 1$.  Figure \ref{fig7}a shows that AD reduces the ion-neutral linewidth ratio below unity; for $\radlo=0.12$, which is close to the hydrodynamic limit, we find $\sigma_{v,i}/\sigma_{v,n}=0.62\pm0.01$.  Figure \ref{fig7}b shows that this result remains valid when the density-weighted velocity dispersions are used instead; in this case, $\sigma_{v,i}/\sigma_{v,n}=0.60\pm0.01$ for $\radlo=0.12$.  However, observers are not able to obtain 3D velocity dispersions as in the block decomposition method.  
Therefore, we also plot the ratio $\sigma_{v,i,\,\rm proj}/\sigma_{v,n,\,\rm proj}$ of the density-weighted velocities normal to the mean field projected through the whole box in Figure \ref{fig7}b (solid symbols).
The ratios are the mean values along the two cardinal directions normal to the mean field.
The ratios from the projected 
$\sigma_{v,i}$ and $\sigma_{v,n}$ show the same behavior as a function of $\rad$ as the ratios from the block decomposition method.  It should be noted, however, that the situation is quite different for observations parallel
to the mean field; in that case, the ion-neutral linewidth ratio is about unity.
Comparison of Figure 3 in \citet{hou02} with Figure \ref{fig7}b then suggests that most of the molecular clouds they observed correspond to $\radlo \ga 12$, which is consistent with the logarithmic mean
$\radlo \sim 20$
for observed molecular clumps that we found in Paper II.  \citet{hou02} found some clouds with ion/neutral velocity-dispersion ratios considerably below the average, and it is possible that these indicate regions of strong AD.

As noted above, \citet{lih08} suggested that the ion-neutral line width ratio is less than unity because the ion velocity dispersion is damped below the AD length scale whereas the neutral velocity dispersion is not.   
\citet{fal10} presented results that support this
idea, and \citet{til10} presented direct two-fluid simulations that demonstrate its validity.
Our higher resolution results are consistent with the work of \citet{til10}.  
However, in contrast with the suggestion of \citet{lih08}, AD is unlikely to produce a sharp break in the ion velocity power spectrum. Even if we assume that the turbulence satisfies a line-width size relation and that the gas is in ionization equilibrium, $\rad$ is still a function of three free parameters (density, velocity, and magnetic field strength), all of which vary in a turbulent medium.  Therefore, instead of finding a sharp change in the slope of the ion velocity power spectrum, we expect a gradual change from an Iroshnikov-Kraichnan spectrum to one that is close to a Burgers spectrum over $\sim 2$ orders of magnitude of $\radlo$, as shown in Figure \ref{fig2}.  None of the AD simulations of \citet{ois06}, \citet{dow09}, or \citet{til10}, our $256^3$ (Paper I) or $512^3$ simulations reveal a change in spectral slope in the inertial range of the calculation. One would need an extremely high resolution simulation with an inertial range spanning $2\sim3$ orders of magnitude of length scale to observe the gradual change in the ion spectrum.

\section{The Effect of Ambipolar Diffusion on the Chandrasekhar-Fermi Method}

Polarimetry is a commonly used method to determine the magnetic field orientation in the plane of the sky \citep[e.g][]{hil00,nov00,war00}.  Based on a dynamical method proposed by \citet{cha53}, the magnetic field strength in the plane of the sky can be estimated provided that the turbulence is isotropic and that the perturbed magnetic energy is in equipartition with the turbulent kinetic energy.  If one can also measure the magnetic field along the line-of-sight using Zeeman mapping, one can obtain the true magnetic field strength and the 3D orientation of the magnetic field.  

\citet{hei01} have carried out a systematic investigation of the accuracy of the Chandrasekhar-Fermi (CF) method using 3D simulations of ideal MHD turbulence.  They proposed a new recipe for using the CF method to estimate the mean magnetic field strength that allows for non-equipartition of turbulent magnetic and kinetic energy, large-amplitude fluctuations of the magnetic field, and the limitations of observational resolution.
They define the ratio of the field inferred via the CF method to the actual field to be 
\beq
a_{\rm CF} \equiv \frac{B_{\rm CF}}{\langle B \rangle},
\eeq
and conclude that, for strong fields at least, $a_{\rm CF} \simeq 2$ when the angular structure of the field is well resolved.

Heitsch et al.'s (2001) result for $\acf$ allows for deviations from equipartition between the energy in nonthermal motions and that in fluctuating magnetic fields, which they parametrize by
\beq
\xi\equiv\frac{\delta U_B}{E_K}=\left(\frac{\avg{\delta B^2}}{\avg{B^2}}\right)\frac{1}{\ma^2},
\eeq
where $\delta U_B$ is the energy in the fluctuating magnetic field and $E_K$ is the
energy in turbulent motions.
Their analysis of the CF method  is based on the assumption that $\xi$ is known.  
The value of $\xi$ has a significant effect on the strength of the inferred field: small values of $\xi$ lead to small fluctuations in the direction of the field; if it is then assumed that $\xi=1$, as in standard applications of the CF method, the smallness of the fluctuations is interpreted being due to a strong field, with
$a_{\rm CF}=\xi^{-1/2}$ \citep{hei01}.
Both observation and theory are consistent with approximate equipartition ($\xi\sim 1$).  Although \citet{hei01} suggested that observations by \citet{cru99} do not support equipartition, the median \alfven\ Mach number of the 15 clouds in Crutcher's sample with measured magnetic fields is unity, which is consistent with $\xi\sim 1$ since $\avg{\delta B^2}\sim \avg{B^2}$ for $\ma\sim 1$. 
(It should be noted, however, that this simple argument allows $\xi$ to deviate
from unity by a factor $\sim$ 2.)  Theoretically, it has been shown that equipartition applies to weak \alfvenic\ turbulence \citep{zwe95}, including that modeled in the theory of \citet{gol97}.  However, there is good evidence for deviations from equipartition in simulations.
\citet{hei01} were agnostic as to whether these deviations are a physical effect or an artifact of their simulations.  One factor that appears to affect the
value of $\xi$ is the driving scale: Simulations in which the driving is on large scales and the field is strong ($\beta\la 0.2$)
(\citealp[e.g.][]{hei01,ost01,lem09} and this work) have $\xi\la$ a few tenths; for example,
we find $\xi\sim 0.2$ in our ideal MHD turbulence model, m3i.
On the other hand, the simulations of \citet{sto98}, in which the driving was on a scale small compared to the box scale, found $\xi\sim 0.6$
for $\beta\la 0.1$. This result can be understood as being due to approximate equipartition between the magnetic fluctuations and the turbulent motions normal to the field lines;
for isotropic turbulence, this would lead to $\xi\simeq \frac 23$. 
In our simulations, which are driven at large scales,
we have found that the reduction in $\xi$ appears to
be due at least in part to motions in which the gas rotates around the field lines. While this may occur in simulations with periodic boundary conditions, 
it is unlikely to occur in Nature.   

Here we address a separate issue: How does AD affect the CF method?
In Paper I, we showed that AD reduces the turbulent magnetic field energy below that in ideal MHD turbulence (see Figure 7 in that paper). 
That is, in addition to the physical and numerical effects that cause $\xi$ to differ from unity in the ideal case, AD reduces $\xi$ because the ions are no longer as well coupled to the dominant neutrals.  Whereas \citet{hei01} studied a range of field strengths, all our models have relatively strong fields, with a plasma-$\beta$ parameter $\beta=0.1$. The level of turbulence is modest, with $\calm=3$ and $\ma=0.67$.

Our treatment of the CF method follows the procedures given in \citet{hei01}, and we use similar notation.  
In terms of the Stokes parameters $U$ and $Q$,
the polarization angle $\phi$ is given by
\beq
\phi = \frac{1}{2} {\rm arctan} \frac{U}{Q},
\eeq
When viewed along the $y$-direction (recall that the mean field is in the $z$-direction), 
the polarized intensity is
\beq
P = Q + iU \propto \int \rho(y) \frac{(B_x+iB_z)^2}{B_x^2+B_z^2}\cos^2\gamma \;dy,
\eeq
where $\gamma$ is the angle between the magnetic field and the plane of the sky.
The CF estimate of the mean field is
\beq
B_{\rm CF} = \sqrt{4 \pi \avg{\rho}} \frac{\sigma(v_{\rm los})}{\sigma(\tan\delta)} \equiv a_{\rm CF}\langle B \rangle
\eeq
where $\avg{\rho}$ is the mean density, $\delta\equiv (\phi-\avg{\phi})$ is the difference between the local and the mean polarization angle, and $\sigma$ is the standard deviation. 
Note that all quantities, including the line-of-sight velocity dispersion, $\sigma(v_{\rm los}),$ are density-weighted.  Because the turbulence is relatively weak in our simulations ($\ma=0.67$), the perturbed field is less than
the mean field.  To maximize the effect of AD, we observe the data perpendicular to the mean-field direction, computing $U$ and $Q$ from $B_x$ and $B_z$ when observed in the $y$-direction (and $B_y$ and $B_z$ when observed in the $x$-direction). We take $B_{\rm CF}$ to be the mean of the estimated field strengths along the $x$ and $y$ directions.

In the presence of AD, the departure from equipartition will be larger than in the ideal case because of the inefficient transfer of turbulent kinetic energy to magnetic energy, as
was demonstrated analytically by \citet{zwe95} and numerically in Paper I (see Figure 7).  In order to focus on the effect of AD without the complications introduced by deviations from equipartition in the ideal MHD case, we normalize the
values of $a_{\rm CF}$ from our simulations by the value of $a_{\rm CF}$ from the ideal MHD model m3i.
(For this model, $\acf\simeq 2$, similar to the value found by \citet{hei01}.) 
The results in Figure \ref{fig8} show that AD starts to affect the CF method somewhere in the range $\radlo = 1 \sim 10$.  At $\radlo \sim 1$, we find that $a_{\rm CF} \sim 3$ times the ideal MHD value. In other words, if we assume that our estimate of the CF coefficient, $\acf$, is correct for the ideal case, then application of the CF method to a system with $\radlo\simeq 1$ would give a field strength that is too large by a factor 3.  
At $\radlo=0.12$, the normalized CF coefficient is $a_{\rm CF}>17$.  Thus, a large correction to the CF method will be required if AD is very strong.
\citet{zwe90} also found that the magnetic field strength would be overestimated in the presence of AD using a highly simplified clump and interclump gas model.
However, the 15 molecular clouds with measured magnetic fields discussed in Paper II have a logarithmic mean $\radlo \sim 20$, and most of them have $\radlo > 10$.  According to Figure \ref{fig8}, AD has only a minimal effect on the inferred field strength, $B_{\rm CF}$, for such clouds, and it is safe to estimate the mean B field based on ideal MHD models.  For regions where $\radlo \la 1$, a significant correction for AD would be necessary.  

\section{Turbulent Enhancement of Ambipolar Diffusion}

Enhancement of diffusion processes by turbulence is observed in experiment, observations, and numerical simulations \citep[e.g.][]{cow09,liu09,nag09}.  
The enhancement of AD by turbulence has been discussed by \citet{zwe02},
\citet{fat02}, \citet{kim02}, and \citet{hei04}.  Each of these works made substantial
simplifications: \citet{fat02} considered field fluctuations in one dimension, 
and the remaining authors assumed that the motions were confined to two dimensions; in addition, \citet{kim02} assumed that the gas was incompressible. Here, we shall present the first 
fully three-dimensional
treatment of the effects of turbulence on AD.

There are two distinct effects of turbulence in the case of AD: First, turbulence increases the
AD drift velocity, $v_d$, and second, it mixes regions on finer and finer scales, so
that less diffusion is needed to smooth out variations in the mass-to-flux ratio.
The first effect determines the rate of AD in a turbulent medium and can be evaluated
for steadily driven turbulence; that is the objective of this section. The second is intrinsically
time dependent, and depends on the initial conditions; we do not address it here,
although the effects of turbulent mixing are naturally included in our simulations.
The combination of the two effects is turbulent AD, and \citet{zwe02} and \citet{kim02} have shown that when the neutrals and ions are well coupled (large $\rad$), 
the diffusion rate for turbulent AD is of order the classical turbulent value,
$v\ell_0$, where $v$ is the turbulent velocity on the scale $\ell_0$.

In molecular clouds, the ionization is generally so low that the ion inertia is negligible. 
In that case, the relative velocity between the ions and neutrals is determined by
the balance between the Lorentz force on the ions and the drag due to the neutrals,
\beqa
\gad\rho_i\rho_n v_d&=&\frac{1}{4\pi}\left|(\vecnabla\cross\vecB)\cross\vecB\right|,\\
&\equiv&\frac{\brms^2}{4\pi\ell_B},
\label{eq:ellb}
\eeqa
which defines $\ell_B$.
Note that this relation is exact in the limit in which the ion inertia is negligible, and the
definition of $\ell_B$ here differs from the approximate expression in previous papers in
this series. In terms of the diffusivity
\beq
\lambda\equiv \frac{\brms^2}{4\pi\gad\rho_i\rho_n}
\eeq
the drift velocity is
\beq
v_d=\frac{\lambda}{\ell_B}.
\label{eq:vd}
\eeq
Equation (\ref{eq:vd}) shows that the effects of turbulence on the
drift velocity can be divided into two parts, the effect on the diffusivity, $\lambda$, and
the effect on the length scale of the Lorentz force, $\ell_B$.
The study by \citet{fat02} focused on the diffusivity: they
pointed out that fluctuations in the density and magnetic field in a turbulent system can increase $\lambda$ and thereby enhance the AD rate.  
Their studies suggest that turbulent fluctuations can enhance the AD rate by a factor $\Lambda \sim 1-10$; for molecular clouds, they estimated $\Lambda \sim 2-3$.  \citet{zwe02}, \citet{kim02} and \citet{hei04} considered both the effect of turbulence in reducing $\ell_B$ and the effects of turbulent mixing, albeit in 2D. They found that the turbulent AD time is of order the eddy turnover time for large $\rad$, which corresponds to a very large enhancement. 
\citet{zwe02} argued that this large enhancement could explain the observed flatness of the relation between magnetic field strength and number density (the $B-n$ relation) in interstellar clouds. 

With the simulations we have performed, we can address the issue of how turbulence
affects the rate of ambipolar diffusion through its effects on the drift velocity. Our results include both the effects of turbulence on the diffusivity, $\lambda$ and on the length scale, $\ell_B$.  A direct measurement of the amount of AD occurring in the turbulent box is the value of of the mass flux relative to the magnetic field, $\rho_n v_d\simeq \rho v_d$, averaged over the volume of the box. 
In order to determine the enhancement factor due to turbulence, we must have a non-turbulent reference value. For large $\rad$, we define the reference value of the drift velocity, $v_{d0}$, as the value of the drift velocity in a uniform medium in which the field varies on the box size, $\ell_0$:
\beq
v_{d0}=\frac{\brms^2}{4\pi\gad\bar\rho_i\bar\rho_n\ell_0}=\frac{v}{\radlo}.
\eeq
Note that we have used the rms field strength here so as to exclude 
from the enhancement factor the effects
of the field amplification
that occurs in a high-$\beta$ plasma.
For small values of $\radlo$, the drift velocity cannot exceed the rms velocity, $v$, so
we generalize the expression for the reference velocity to
\beq
v_{d0}=\frac{v}{1+\radlo}.
\eeq
The enhancement factor for AD due to turbulence is then
\beq
\Lambda=\frac{\left\langle\rho_n v_d\right\rangle}{\bar\rho_n v_{d0}}=
\frac{1+\radlo}{\radlo}\left\langle\frac{\bar\rho_i\ell_0}{\rho_i \ell_B}\right\rangle.
\label{eq:lambda}
\eeq

Values of the time-averaged turbulence enhancement factor $\Lambda$ are listed in Table 4 for all five AD models.  For $\radlo\geq 12$, the enhancement is a factor of 2 to 4.5.  For $\radlo\leq 1.2$, the enhancement is negligible ($\Lambda\simeq 1$): the ions and neutrals are sufficiently decoupled that turbulence does not have a significant effect on the AD rate (the apparent divergence in $\Lambda$ due to the factor $\radlo$ in the denominator of eq. \ref{eq:lambda} is more than cancelled by the increase in the magnetic scale length, $\ell_B$ --- gradients in the field cannot be maintained in a weakly coupled plasma).  Our results for the turbulent enhancement of AD should apply to non self-gravitating regions in the ISM, as studied by \citet{zwe02}. Our results do not apply to AD in self-gravitating regions, such as star-forming cores. 

\citet{kud08,kud11} have shown that large-scale turbulence accelerates the formation of magnetically supercritical cores (i.e., cores in which gravity dominates the magnetic field) by compressing the gas. The turbulence was initialized in the plane of the slab and allowed to decay. \citet{kud11} found that the time required for bound cores to form scales as the inverse square root of the peak density produced by turbulent compression, prior to the time that gravity was dominant. Since the ambipolar diffusion time and the gravitational free-fall time both scale as $\rho^{-1/2}$ for the typical condition in which the ionization also scales as $\rho^{-1/2}$, this is just what is expected for a compression in the absence of turbulence (e.g., \citealp{mou99}). The simulations of \citet{kud08,kud11} did not address the effect of turbulence on the AD time within individual cores, in part because the turbulence damped with time. 
It is possible that the turbulent enhancement factor inside the cores would be reduced by the effects of self-gravity since flux tubes are then not as free to 
move, but confirming this must await a future study.

\section{Discussion and Conclusions}

\begin{figure}
\epsscale{.80}
\includegraphics[scale=0.7]{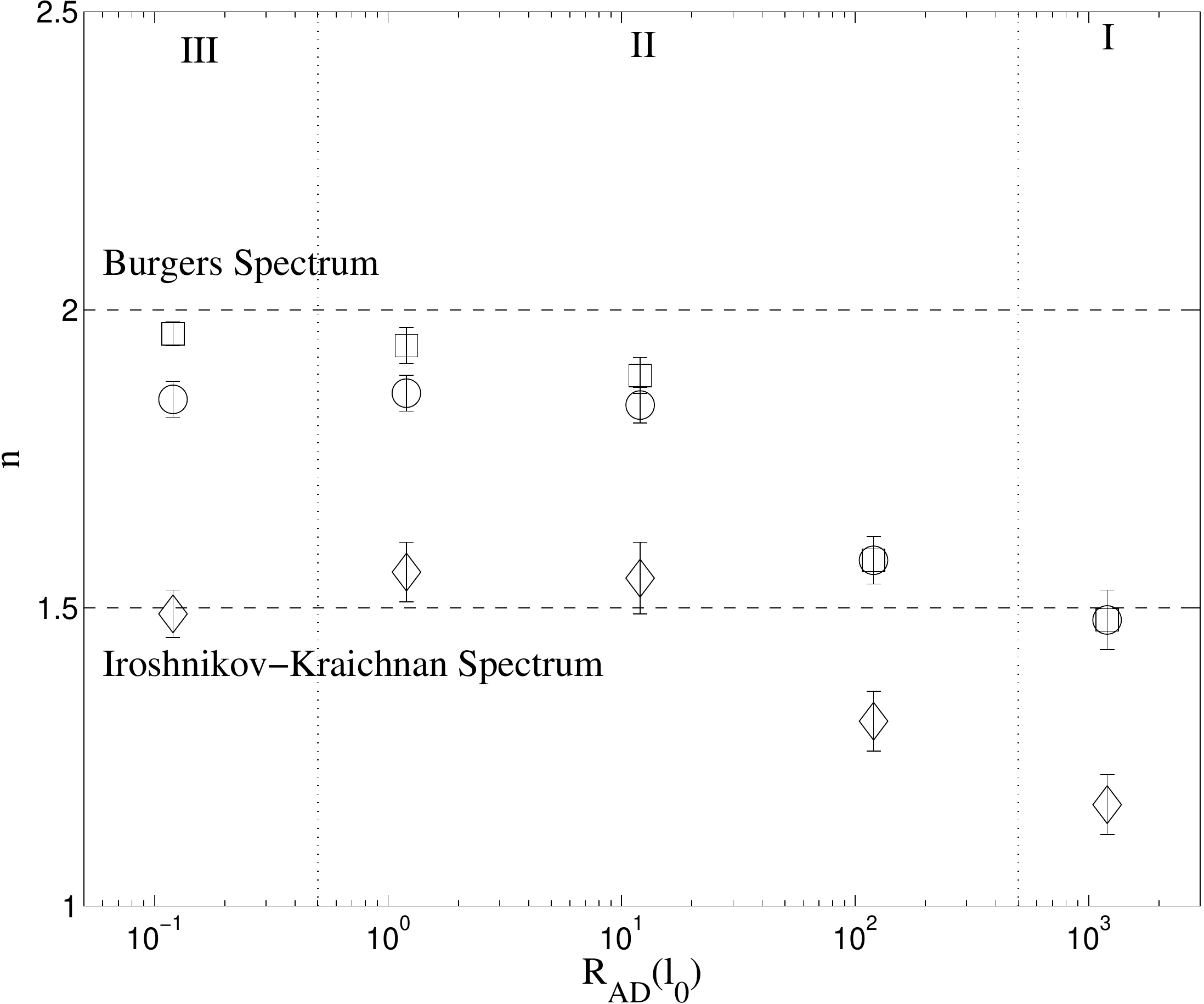}
\caption{Velocity and magnetic field power spectral indexes of five AD models.  The neutral velocity power spectral index (squares) changes from an Iroshnikov-Kraichnan spectrum to a Burgers spectrum in going from AD regime I to regime III.  The ion velocity spectral index (circles) also increases as AD becomes important, but it diverges from the neutral velocity index and stays at about $n=1.85$ as the ions decouple from the neutrals.  The magnetic field power spectral index (diamonds) has a similar trend as $\rad$ decreases.
\label{fig2}}
\end{figure}

\begin{figure}
\epsscale{.80}
\includegraphics[scale=0.7]{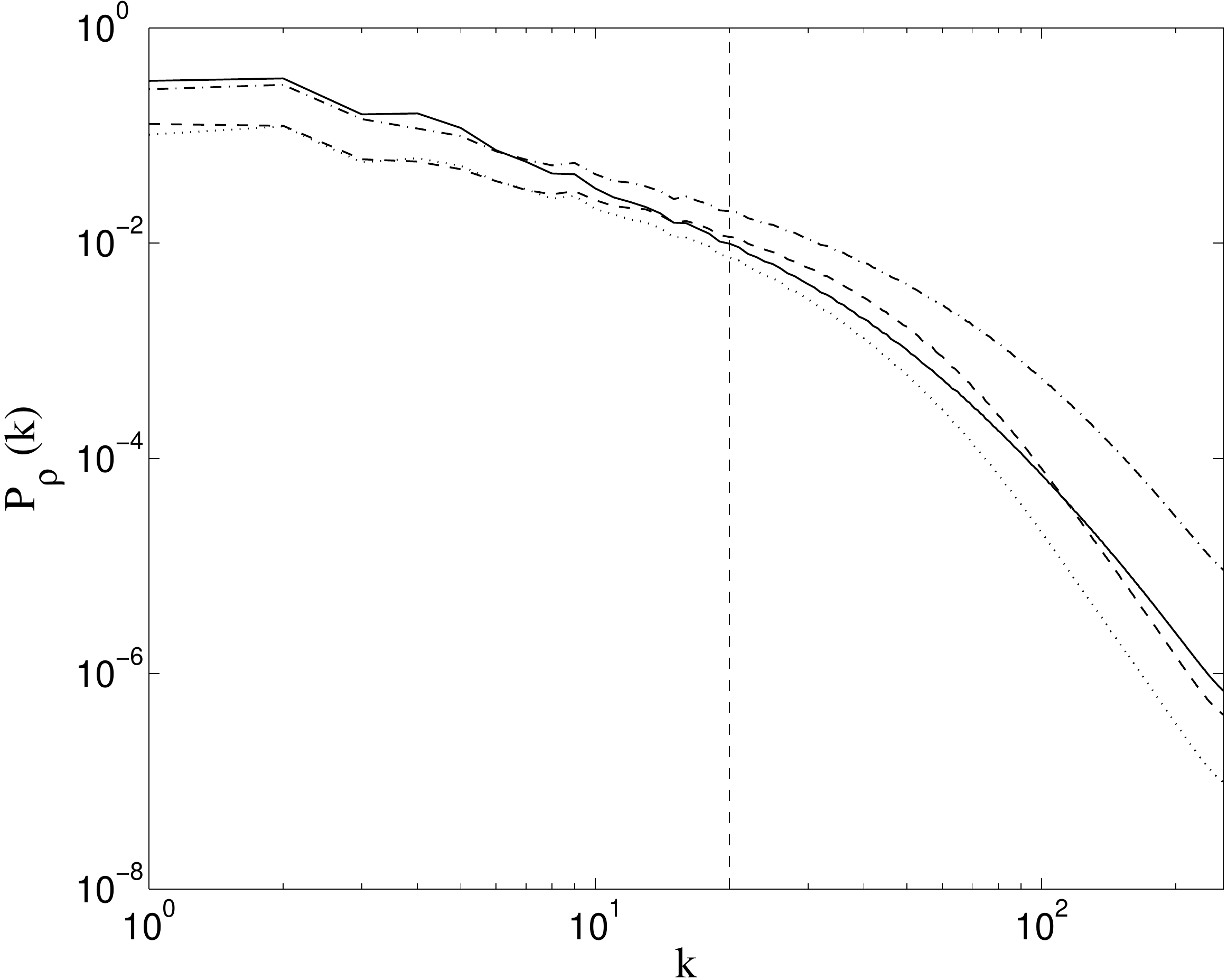}
\caption{
Time-averaged
density power spectrum for ideal MHD (dashed; model m3i) and neutral density power spectra for $\radlo=120$ (dotted; m3c2r2), $\radlo=12$  (solid; model m3c2r1), and $\radlo=0.12$  (dot-dashed; model m3c2r-1).  The spectral index $n_\rho$ steepens to a maximum as $\radlo$ decreases from infinity (ideal MHD) to $\radlo = 12$ and becomes shallower again as $\radlo$ decreases further.  See \S\ref{sec:power} for discussion of the relation between the change in the morphology of molecular clouds and the density power spectrum.
\label{fig3}}
\end{figure}

\begin{figure}
\epsscale{.40}
\plotone{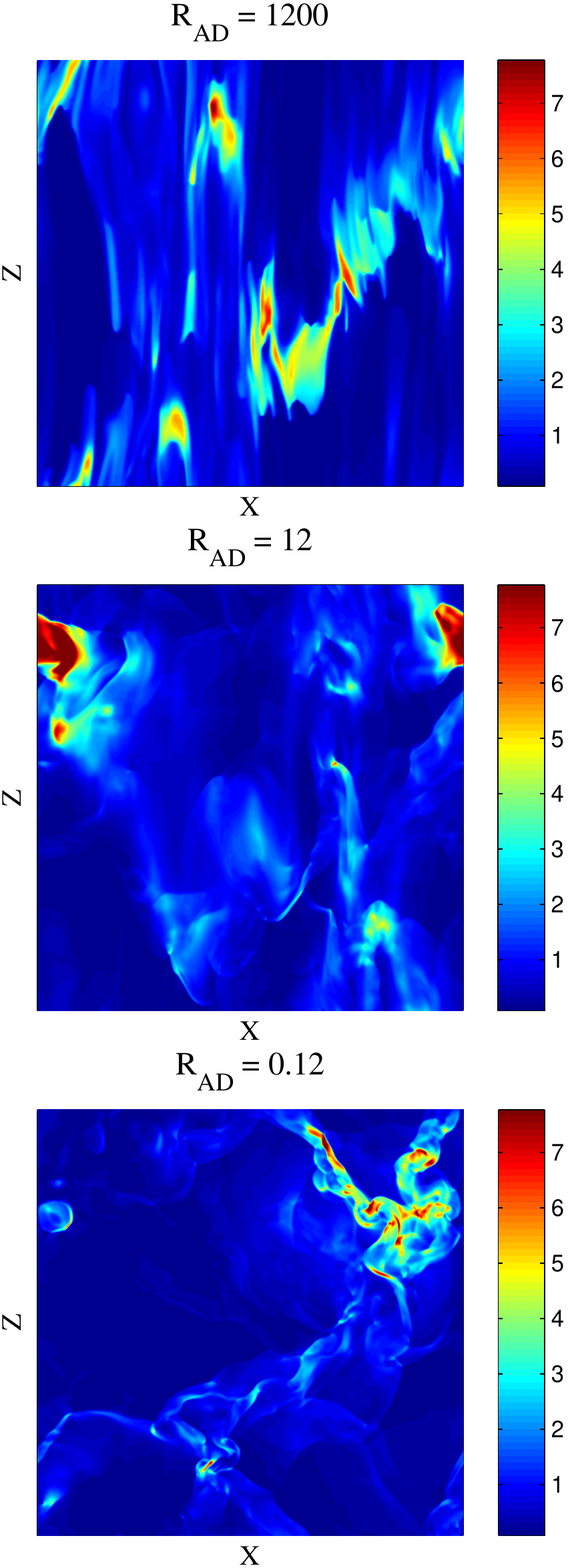}
\caption{Density slices from models (a) m3c2r3 ($\radlo=1200$), (b) m3c2r1
($\radlo=12$), and (c) m3c2r-1 ($\radlo=0.12$), showing the morphological changes in the density distribution as the effect of AD becomes stronger.  Panel (a) shows sheet and filamentary distributions mainly along the mean direction of the magnetic field.  Panel (b) shows diffusion of the sheet and filamentary distributions.  In panel (c), many high-density small structures form as the result of fragmentation.  See \S\ref{sec:power} for discussion of the relation between the change in the morphology of molecular clouds and the density power spectrum.
\label{fig4}}
\end{figure}

\begin{figure}
\epsscale{1}
\plotone{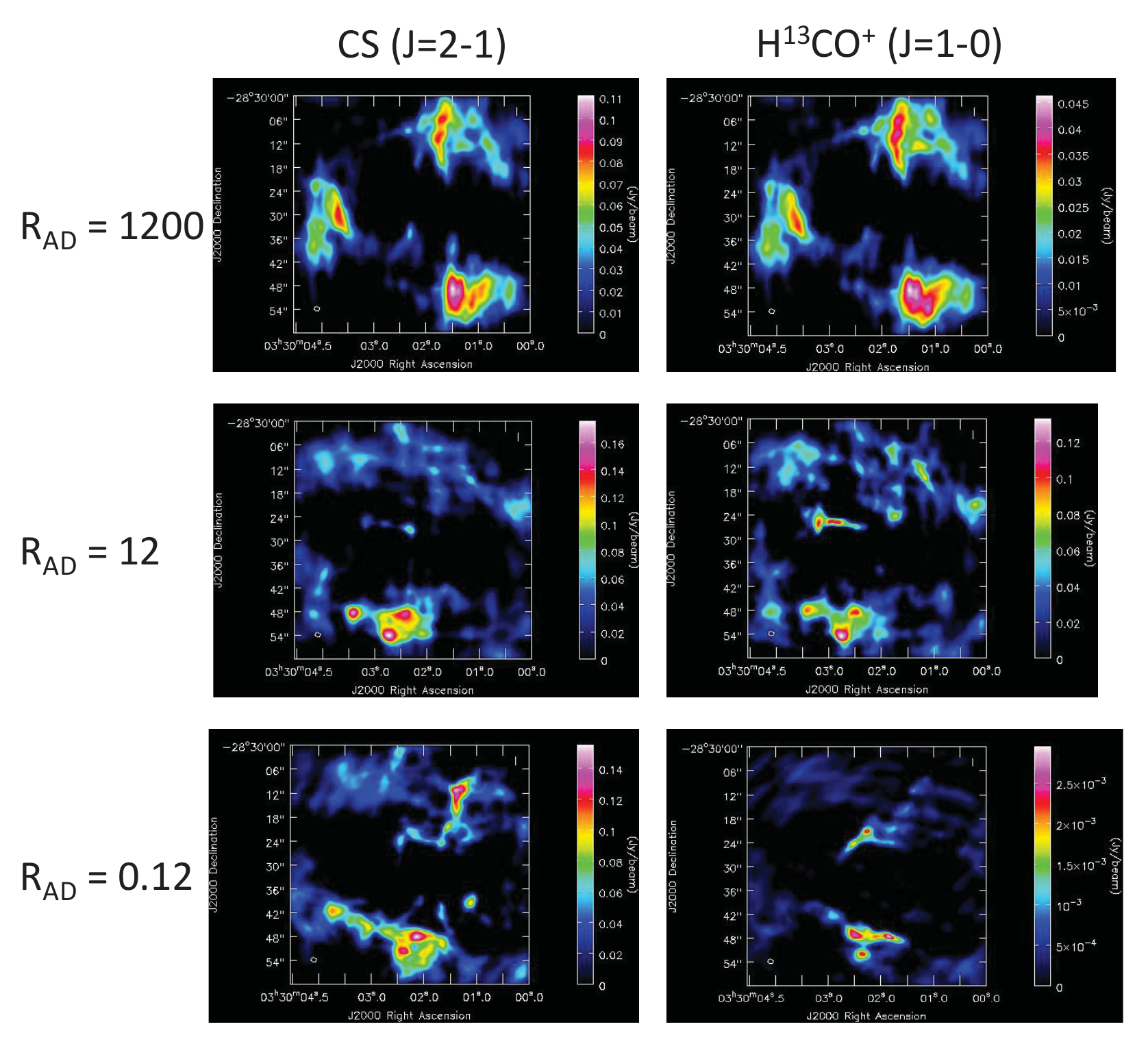}
\caption{
Simulated sky images, created from simulation data and post-processed with the radiative transfer code SimLine, that would be observed with the ALMA telescope.  This figure shows simulated images of the ions (deduced from the H$^{13}$CO$^+$ J = 1-0 transition---left column) and neutrals (deduced from the CS J=2-1 transition---right column) in a molecular cloud 
with a density of $10^4$ cm\eee\ 
as observed by ALMA for 2 hours.
The top panels are from model m3c2r3 ($\radlo = 1200$), the middle panels are from model m3c2r1 ($\radlo = 12$), and the bottom panels are from model m3c2r-1 ($\radlo = 0.12$).  With very weak AD (top row), the images of the ions and the neutrals will appear almost the same (\S\ref{sec:power}).  The shapes and sizes of the beams are shown at the left bottom of the images. (Color images available in the electronic version.)
\label{fig5}}
\end{figure}

\begin{figure}
\epsscale{.80}
\includegraphics[scale=0.7]{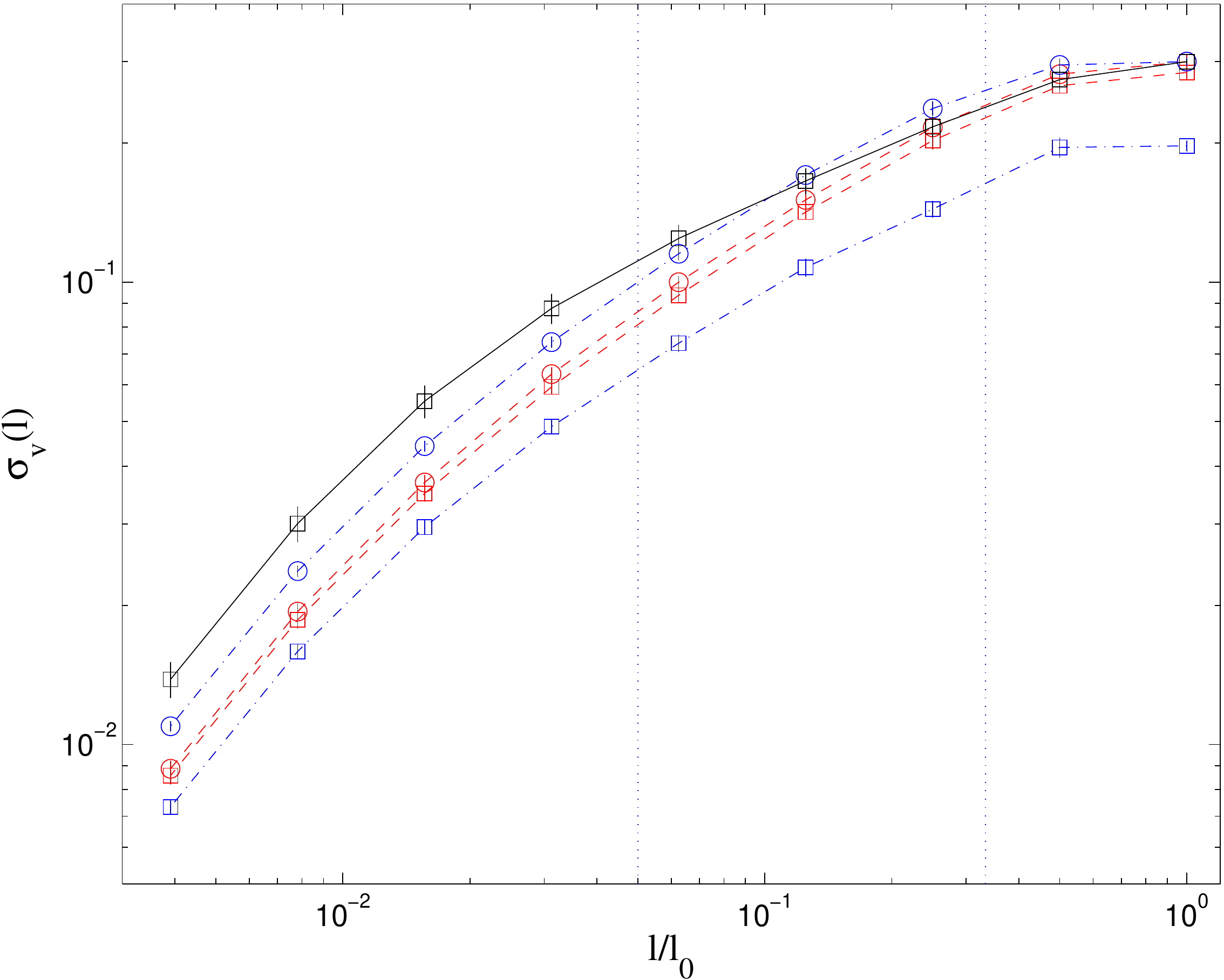}
\caption{
Time-averaged
linewidth-size relation for models m3c2r-1 (dot-dashed; $\radlo=0.12$), m3c2r1 (dashed; $\radlo=12$), and  model m3i (solid; ideal MHD).  The ion linewidth-size relation is marked by squares and the neutral linewidth-size relation is marked by circles.
The velocity dispersions are computed using the box decomposition method.
The linewidth-size relations obey a power law in the inertial range (the region between the two dotted vertical lines) from the power spectra.  See \S\ref{sec:lws} for a discussion of the linewidth-size relation for different models.  The power-law indexes of the linewidth-size relation for each model are listed in Table 3.
\label{fig6}}
\end{figure}

\begin{figure}
\includegraphics[scale=0.7]{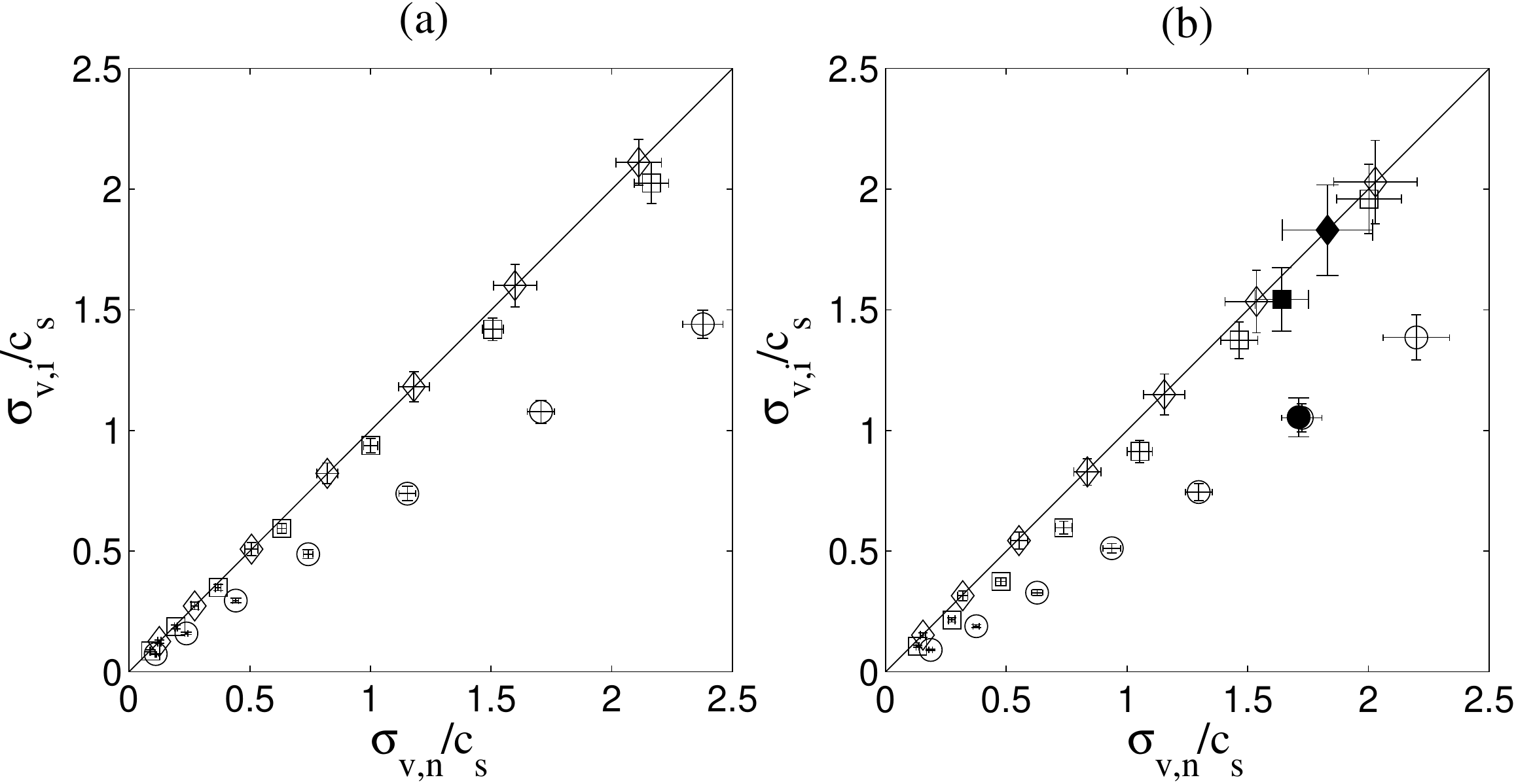}
\caption{Plots of ion and neutral velocity dispersions, $\sigma_{v,i}$ and $\sigma_{v,n}$ respectively, for models m3c2r-1 (circles), m3c2r1 (squares), and m3c2r3 (diamonds). (a) Volume-weighted mean $\sigma_{v,i}$ and $\sigma_{v,n}$ obtained from the box decomposition method. (b) Density-weighted mean $\sigma_{v,i}$ and $\sigma_{v,n}$ obtained from the box decomposition method. The solid line corresponds to $\sigma_{v,i}/\sigma_{v,n} = 1$.  The 3 solid symbols are the velocity dispersions of the corresponding models projected normal to the mean field.  See \S\ref{sec:inlr} for discussion. 
\label{fig7}}
\end{figure}

\begin{figure}
\epsscale{.80}
\includegraphics[scale=0.7]{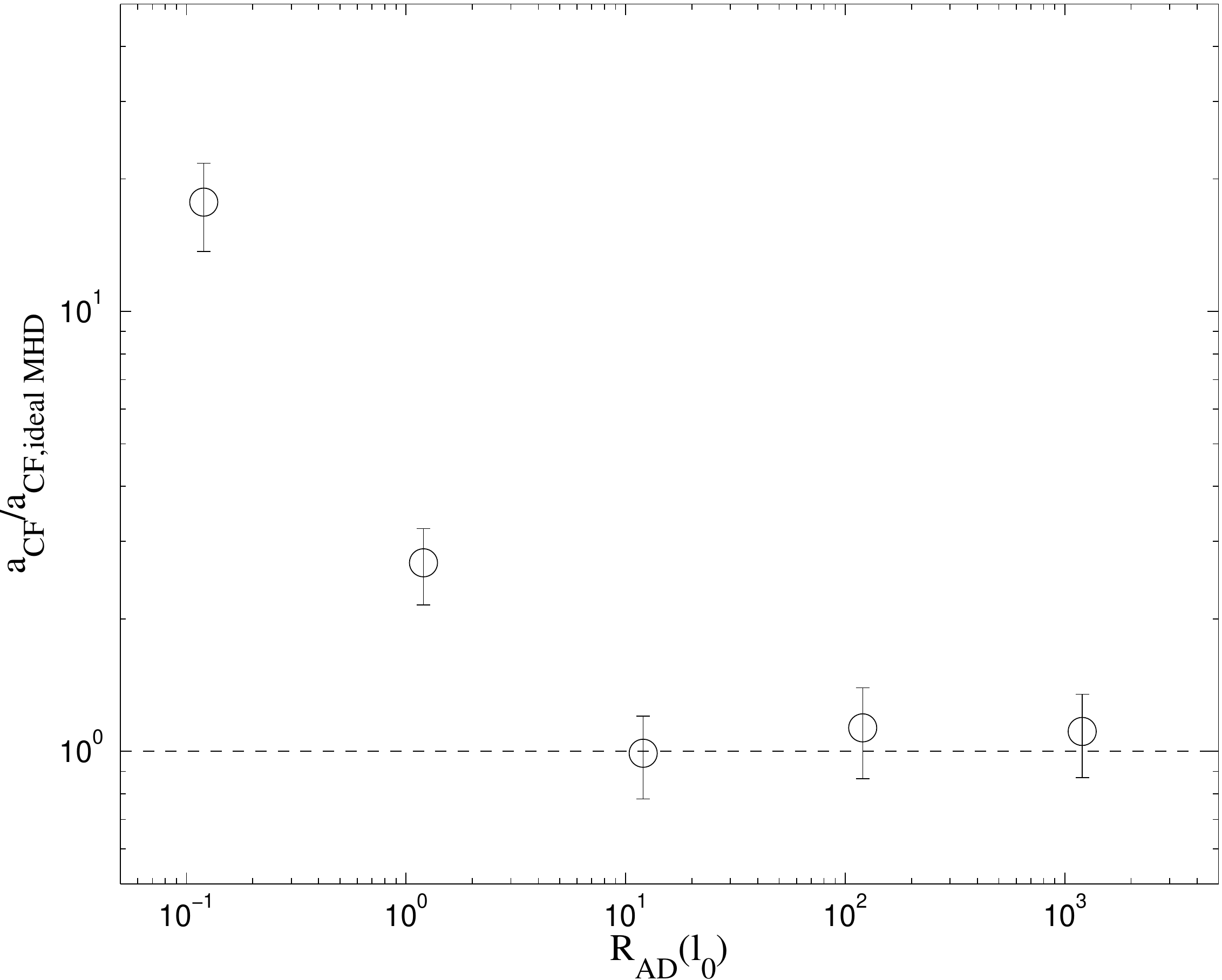}
\caption{Normalized correction factor $a_{\rm CF} = B_{\rm CF}/\langle B \rangle$, which is the ratio of the estimated magnetic field strength using the CF method to the true magnetic field strength.  The correction factor is normalized to the correction factor from the ideal MHD model m3i to show the effect of AD on the CF method.  The error bars show the uncertainty of the time-averaged mean of the measurements from all the data dumps over two crossing times. For $\radlo \ga 10$, AD  has a negligible effect on the inferred magnetic field strength.  AD starts to have an effect on the correction factor when $1 < \radlo < 10$, and the correction factor is $\sim 3$ at $\radlo = 1$.  For $\radlo < 1$, the field strength would be significantly overestimated using the CF method.
\label{fig8}}
\end{figure}

In this paper, we have investigated a series of issues on the observational implications of molecular clouds from the effects of turbulent AD.  We summarize our findings as follows:

\begin{itemize}
\item[1.] We confirm the results from the $256^3$ models in Paper I that the velocity and magnetic power spectra are functions of the AD Reynolds number, $\rad$.  
The higher resolution $512^3$ models reported in this paper show that the transition of the neutral velocity spectrum from the Iroshnikov-Kraichnan spectrum for a strong field, ideal MHD model to the Burgers spectrum for a shock-dominated, pure hydrodynamic model occurs primarily in the Standard AD regime (Fig. \ref{fig2}).  
The ion velocity spectrum diverges from the neutral power spectrum when the coupling between ions and neutrals becomes weak, for $\radlo \la 10$.  The magnetic field power spectrum shows similar behavior.  The 15 molecular clumps with measured magnetic fields \citep{cru99} are all inside the Standard AD regime, and 
their velocity power spectra should show this transition when the high-$\radlo$ clumps are compared with the low-$\radlo$ ones (see paper II).

\item[2.] The density power spectrum steepens as AD becomes important and reaches a maximum at $\radlo \sim 10$; it then becomes shallower as the neutrals decouple from the ions at still lower values of $\radlo$.  The steepening occurs because AD is most effective on small scales, which leads to damping of the density fluctuations on these scales.  When the neutrals decouple from the ions, neutral gas clumps fragment and the spectrum approaches the hydrodynamic limit.  The change in the density spectral index corresponds to the change in the morphology of the gas.  We have applied the CASA software simdata2 to determine the intensity of emission in the CS $J=2-1$ and
HCO$^+\; J=1-0$ transitions implied by our simulations. The results
demonstrate that the decoupling of ions and neutrals is observable in principle.

\item[3.] We have investigated the linewidth-size relation from our simulation models using a 3D box decomposition method.  This method distorts the physical linewidth-size relation because the power in a box is reduced by numerical damping at small scales.  The best way to calculate the linewidth-size relation in the inertial range is to use equation \ref{eq:q} to infer the exponent in the linewidth size relation from the power-law index of the velocity power spectrum.

\item[4.] Our high-resolution two-fluid simulations confirm the suggestion of \citet{hou00a} that the reduction of the observed ion-neutral velocity dispersion ratio to values below unity is the result of AD.  However, our results do not support the microphysical model of \citet{lih08} in which the damping of the ion motions sets in at a well-defined length scale; instead, we find that the effects of AD become increasingly important over $\sim 2$ orders of magnitude in length scale.
It should be noted that our isothermal simulations do not include drift heating of the ions, and therefore provide lower limits on the actual ratio of the ion-neutral velocity dispersion ratio when observed normal to the field; on the other hand, this field orientation is the optimum one for observing the effect, and we find that the ion-neutral linewidth ratio is about unity when observed along the field.  Our results also show that most of the observed ion-neutral linewidth ratios in the \citet{li08} sample of molecular clouds correspond to $\rad \ga 10$, consistent with the computed $\rad$ of the 15 molecular clumps discussed in Paper II.

\item[5.] We have examined the effect of AD on the Chandrasekhar-Fermi method, which is commonly used to estimate the magnetic field strength in the plane of the sky from observations of polarization angles and velocity dispersions.  When AD is significant, the magnetic field will be less perturbed and, therefore, the corresponding mean field strength will be over-estimated.  Our analysis shows that AD does not significantly affect the CF estimate of the field in the range of $\rad$ of the molecular clumps in Crutcher's (1999) sample,  
There is no need to correct values of the magnetic field estimated using the CF method for the effects of AD except in the very small regions 
in which $\radlo \le 1$.

\item[6.] With the heavy-ion approximation, we have been able to carry out the first fully three-dimensional study of the enhancement of AD by turbulence.  Our simulations show that when the ions and neutrals are reasonably well coupled ($\radlo\geq 12$, as is the case for most of the molecular clumps analyzed in Paper II), AD in non self-gravitating regions in the ISM is accelerated by a factor $\sim 2 - 4.5$ for the conditions we consider.  
In the weak ion-neutral coupling regime, AD is so strong that turbulence cannot significantly enhance the rate of AD.

\end{itemize}
\noindent

\acknowledgments
We would like to thank S.-P. Lai, F. Heitsch, and E. Zweibel for discussions of the CF method.  We also thank Volker Ossenkopf on the suggestions of using the 1D raditation transfer code SimLine and the referee, R. Banerjee, for his helpful questions and suggestions on improving the manuscript.  Support for this research was provided by NASA through NASA ATP grants NNX09AK31G (RIK, CFM, and PSL), the US Department of Energy at the Lawrence Livermore National Laboratory under contract DE-AC52-07NA 27344 (RIK), and the NSF through grant AST-0908553(CFM and RIK).  This research was also supported by grants of high performance computing resources from the National Center of Supercomputing Application through grant TG-MCA00N020.

\clearpage

\clearpage

\clearpage

\clearpage

\clearpage

\clearpage

\clearpage

\clearpage

\begin{table}
\begin{center}
\caption{Model Parameters and Regimes of AD \label{tbl-1}}
\begin{tabular}{lcccc}
\\
\tableline\tableline
Model$^a$ & $\gad$ & $\rad(\ell_0)$ & Regime of AD \\
\tableline
m3c2r-1 &4        &0.12     &III\\
m3c2r0  &40       &1.2      &II $\sim$ III\\
m3c2r1  &400      &12       &II\\
m3c2r2  &4000     &120      &II\\
m3c2r3  &40000    &1200     &I\\
m3i  	&$\infty$ &$\infty$ &I\\
\tableline
\end{tabular}
\end{center}
$^a$ Models are labeled as ``mxcyrn," where $x$ is the thermal Mach number, $y=|\log\chio|$, and $n=\log(\radl/1.2)$. Model ``m3i" is an ideal MHD. Model m3c2r0 is the same as model m3c2h in paper I.
\end{table}
\clearpage

\clearpage
\begin{table}
\begin{center}
\caption{Spectral indexes of Velocity and Magnetic Field Power Spectra with Varying $\rad$ \label{tbl-2}}
\begin{tabular}{lcccccccc}
\\
\tableline\tableline
                      & m3c2r-1        & m3c2r0        & m3c2r1        & m3c2r2         & m3c2r3          & m3i\\
\tableline
$\rad(\ell_0)$        & 0.12           & 1.2           & 12.0          &
120            & 1200            & $\infty$\\
$n_{\rm vi,r}(k)$     & 1.78$\pm$0.06  & 1.64$\pm$0.06 & 1.80$\pm$0.05 & 1.74$\pm$0.06  & 1.54$\pm$0.05   & 1.48$\pm$0.06\\
$n_{\rm vi,z}(k)$     & 1.89$\pm$0.03  & 1.98$\pm$0.03 & 1.88$\pm$0.04 & 1.40$\pm$0.06  & 1.39$\pm$0.06 & 1.38$\pm$0.05\\
$n_{\rm vi}(k)$       & 1.85$\pm$0.03  & 1.86$\pm$0.03 & 1.84$\pm$0.03 & 1.58$\pm$0.04  & 1.48$\pm$0.05   & 1.45$\pm$0.05\\
$n_{\rm vn,r}(k)$     & 1.97$\pm$0.02  & 1.92$\pm$0.03 & 1.89$\pm$0.04 & 1.76$\pm$0.05  & 1.55$\pm$0.05   & - \\
$n_{\rm vn,z}(k)$     & 1.93$\pm$0.02  & 1.95$\pm$0.03 & 1.88$\pm$0.04 & 1.41$\pm$0.06  & 1.39$\pm$0.06   & - \\
$n_{\rm vn}(k)$       & 1.96$\pm$0.02  & 1.94$\pm$0.03 & 1.89$\pm$0.03 & 1.58$\pm$0.04  & 1.48$\pm$0.05   & - \\
$n_{\rm B,r}(k)$      & 1.43$\pm$0.03  & 1.48$\pm$0.05 & 1.45$\pm$0.06 & 1.23$\pm$0.05  & 1.10$\pm$0.05   & 1.14$\pm$0.07\\
$n_{\rm B,z}(k)$      & 2.07$\pm$0.07  & 2.14$\pm$0.07 & 2.12$\pm$0.05 & 1.88$\pm$0.06  & 1.64$\pm$0.05   & 1.61$\pm$0.08\\
$n_{\rm B}(k)$        & 1.49$\pm$0.04  & 1.56$\pm$0.05 & 1.55$\pm$0.06 & 1.31$\pm$0.05  & 1.17$\pm$0.05   & 1.23$\pm$0.07\\
$n_{\rho{\rm n}}(k)$  & 1.23$\pm$0.03  & 1.44$\pm$0.05 & 1.77$\pm$0.03 & 1.34$\pm$0.03  & 1.08$\pm$0.06   & - \\
$n_{\rho{\rm i}}(k)$  & 0.83$\pm$0.04  & 0.62$\pm$0.04 & 0.68$\pm$0.03 & 0.78$\pm$0.04  & 1.02$\pm$0.05   & 1.08$\pm$0.07\\
\tableline
\end{tabular}
\end{center}
\end{table}

\clearpage
\begin{table}
\begin{center}
\caption{Line Width-Size Relation Power index $q$ Estimated from Different Methods\label{tbl-3}}
\begin{tabular}{lcccccccc}
\\
\tableline\tableline
 & m3c2r-1 & m3c2r0 & m3c2r1 & m3c2r2 & m3c2r3 & m3i \\
\tableline
$\rad(\ell_0)$ & 0.12          & 1.2           & 12            & 120           & 1200          & $\infty$ \\
${q_{\rm i,bd}}^a$ & $0.48\pm0.05$ & $0.51\pm0.03$ & $0.56\pm0.04$ & $0.45\pm0.05$ & $0.42\pm0.05$ & $0.40\pm0.06$ \\
${q_{\rm i,ips}}^b$ & $0.43\pm0.02$ & $0.43\pm0.02$ & $0.42\pm0.02$ & $0.29\pm0.02$ & $0.24\pm0.03$ & $0.23\pm0.03$ \\
${q_{\rm i,dps}}^c$ & $0.55\pm0.03$ & $0.57\pm0.03$ & $0.64\pm0.05$ & $0.52\pm0.03$ & $0.46\pm0.04$ & $0.43\pm0.05$ \\
\tableline
$q_{\rm n,bd}$ & $0.53\pm0.04$ & $0.55\pm0.03$ & $0.56\pm0.03$ & $0.46\pm0.04$ & $0.42\pm0.05$ & - \\
$q_{\rm n,ips}$ & $0.48\pm0.01$ & $0.47\pm0.02$ & $0.45\pm0.02$ & $0.29\pm0.02$ & $0.24\pm0.03$ & - \\
$q_{\rm n,dps}$ & $0.60\pm0.04$ & $0.61\pm0.04$ & $0.63\pm0.04$ & $0.52\pm0.03$ & $0.46\pm0.04$ & - \\
\tableline
\end{tabular}
\end{center}
$^a$ Estimated from the box-decomposition method.

$^b$ Estimated from the inertial range of the ion velocity power spectrum.

$^c$ Estimated from the full range (including dissipation range) of the ion velocity power spectrum.

The subscript notation also applies to the neutral LWS power index $q$ in the second half of the table.
\end{table}

\clearpage
\begin{table}
\begin{center}
\caption{Turbulent Enhancement of the AD Rate. \label{tbl-4}}
\begin{tabular}{lccc}
\\
\tableline\tableline
Model	 & $\rad(\ell_0)$ & $\Lambda$ \\
\tableline
m3c2r-1  & 0.12           & $0.99\pm0.02$ \\
m3c2r0   & 1.2		  & $1.18\pm0.02$ \\
m3c2r1   & 12		  & $2.23\pm0.09$ \\
m3c2r2   & 120		  & $3.11\pm0.24$ \\
m3c2r3   & 1200		  & $4.53\pm0.36$ \\
\tableline
\end{tabular}
\end{center}

\end{table}

\end{document}